\documentclass[prb,aps,twocolumn,showpacs,amsmath,amssymb]{revtex4-1}

\usepackage{times}
\usepackage{graphicx}
\usepackage{epstopdf}
\usepackage{psfrag}
\usepackage{color}
\usepackage{wasysym}
\usepackage[tight]{subfigure}
\definecolor{dred}{rgb}{0.7,0.0,0.0}
\usepackage{dcolumn}% Align table columns on decimal point
\usepackage{bm}% bold math
\allowdisplaybreaks 
\usepackage{braket}
\usepackage{ulem} % to strike things out
\usepackage[]{hyperref}
\hypersetup{colorlinks=true, linkcolor=blue, citecolor=blue, filecolor=blue, urlcolor=blue}

\definecolor{orange}{rgb}{1,0.5,0}
\definecolor{black}{rgb}{0,0,0}

\begin{document}

\title{Two-dimensional topological order of kinetically constrained quantum particles}

\author{Stefanos Kourtis}
\author{Claudio Castelnovo}
\affiliation{TCM group, Cavendish Laboratory, University of Cambridge, Cambridge CB3 0HE, United Kingdom}

\date{\today}

\begin{abstract}
% Motivated by recent experimental and theoretical work on driven optical lattices,
We investigate how imposing kinetic restrictions on quantum particles that would otherwise hop freely on a two-dimensional lattice can lead to topologically ordered states. The kinetically constrained models introduced here are derived as a generalization of strongly interacting particle systems in which hoppings are given by flux-lattice Hamiltonians and may be relevant to optically driven cold-atom systems. After introducing a broad class of models, we focus on particular realizations and show numerically that they exhibit topological order, as witnessed by topological ground-state degeneracies and the quantization of corresponding invariants. These results demonstrate that the correlations responsible for fractional quantum Hall states in lattices can arise in models involving terms other than density-density interactions.
\end{abstract}

\maketitle

\section{Introduction}

The pursuit of topological states of matter has fueled substantial experimental and theoretical developments in physics. From the integer quantum Hall effect~\cite{vonKlitzing1980} to topological insulators~\cite{Hasan2010,Qi2008} and beyond, topological states possess remarkable properties of both fundamental and technological interest. For instance, certain correlated topologically ordered states, such as fractional quantum Hall (FQH) states~\cite{Tsui1982}, have the potential to be used for fault-tolerant quantum computation~\cite{Nayak2008}. The functionality required for novel high-end technological applications, however, steers material design towards microscopic structures with increasingly complex features. 

% Recently,
Several steps have been taken in order to bridge the gap between theoretical descriptions of FQH-type topological order and experimentally accessible systems. Following the first theoretical proposal suggesting the presence of FQH states in a lattice model~\cite{Kliros1991}, FQH physics was predicted to arise in the context of optically trapped cold atoms~\cite{Cooper2001,Cazalilla2005,Cooper2013}. More recently, a class of lattice models called Chern insulators~\cite{Haldane1988} have been shown to develop FQH-like ground states, termed fractional Chern insulators (FCI)~\cite{Neupert2011,Sheng2011,Regnault2011}, upon introduction of a short-range repulsion between particles occupying a fractionally filled band, even in the absence of an external magnetic field (see Refs.~\citenum{Parameswaran2013,Bergholtz2013} for comprehensive reviews). Initially, these models were fine tuned so that the lowest band of their energy dispersion imitates the lowest Landau level: It is almost perfectly flat and topological, i.e., characterized by a Chern number $C=\pm1$ and hence named Chern band. Proposals for realizations of FCI states based on oxide heterostructures~\cite{Xiao}, layered multiorbital systems~\cite{Venderbos2011a,Kourtis2012a}, optical lattices~\cite{Yao2013a}, and strained or irradiated graphene~\cite{Ghaemi2012a,Grushin2013} followed soon thereafter.

As bands in solids are generally not flat and interaction strengths vary, an obvious way to bring FQH-like states closer to reality is to relax the energetic conditions imposed on relevant systems in order to emulate Landau levels. It was recognized early on that interaction strengths may be allowed to increase beyond band gaps~\cite{Hafezi2007,Sterdyniak2012a,Sheng2011,Venderbos2011a}, and subsequent results showed that a finite dispersion may actually favor certain FCI states~\cite{Grushin2012,Kourtis2012a}. One can then venture into the strong-correlation regime by allowing for arbitrarily strong repulsion strengths. Surprisingly, one still finds robust FCI states --- as well as more exotic, topologically ordered states~\cite{Kourtis2013} --- even though interactions now mix bands with opposite Chern numbers~\cite{Sheng2011,Kourtis2013a}. Strongly correlated materials without sharply defined bands may therefore be considered as candidates for the realization of FCI or similar states.

When particles repel their neighbors very strongly, it is reasonable to approximate the interaction as a hardcore constraint which does not allow particles to occupy neighboring sites. On two-dimensional lattices of corner- or side-sharing triangles, the configurations allowed by this constraint are identical to those of the so-called hard-hexagon (HH) model of classical lattice gases~\cite{Runnels1966,Baxter1980}. In the quantum version of these HH models, transitions from one configuration to another are caused by hopping terms. When the hoppings considered give rise to Chern bands, imposing a hardcore constraint in partially filled lattices can lead to FQH-like ground states~\cite{Kourtis2013a}.

Here we explore the possibility of obtaining the same physics by replacing the hardcore constraint with a kinetic one. Instead of energetically penalizing or --- in the infinite-interaction limit --- removing configurations from the Fock space altogether, we start with all configurations being a priori equivalent. Constraints are then introduced only in the transitions between configurations. The resulting systems can be thought of as quantum versions of cooperative lattice gases~\cite{Ritort2002}. Similarly constrained quantum particles have been studied in several contexts in the past, with the kinetic constraints often giving rise to nontrivial correlation effects~\cite{Japaridze1999,Perez2001,Gorski2011,Moller2012}. In this paper, we show that when the constrained kinetic terms are endowed with appropriate Berry phases that break time-reversal symmetry, the obtained ground states show definitive features of FQH topological order.

This result is far from being purely academic: Current experiments on optical lattices can generate a partially filled Chern band in the laboratory, with tunable interaction strengths~\cite{Aidelsburger2014,Jotzu2014}. Some of the terms that arise in the modeling of the experimental setups exhibit precisely the type of constraints studied here~\cite{Aidelsburger2014,Bukov2014}. The same kinetic constraints are also an approximate limit of strong short-range repulsion and therefore relevant to the realization of FQH physics in strongly correlated materials, as will be shown below.

The rest of this paper is structured as follows: In Sec.~\ref{sec:models} we present a general formalism of tightly-bound particles interacting via strong nearest-neighbor repulsion. In the infinite-interaction limit particles obey a hardcore nearest-neighbor exclusion principle, motivating us to re-interpret the problem as a quantum hard-hexagon model. We then relax the static exclusion constraint to a kinetic one, derive two examples of constrained fermionic models, and pose the question: Can topological order arise from the kinetic constraints? In Sec.~\ref{sec:results} we investigate the topological nature of the ground states of the constrained models using exact diagonalization. The results we obtain prompt us to answer the question affirmatively. Our conclusions are summarized in Sec.~\ref{sec:conclusions}.

\section{Models: From strong interactions to kinetic constraints}\label{sec:models}

\subsection{General formalism}

To establish a general formalism, we consider $N$ quantum particles of a single species that hop on a two-dimensional lattice $\Lambda$. The discussion below will be limited to spinless fermions, but there is no fundamental obstacle to treating bosons in the same fashion. Distances between nearest-neighboring sites of $\Lambda$ are set to unity for convenience, and periodic boundary conditions are assumed in both spatial directions. The general form of the Hamiltonians we consider below reads
\begin{equation}
 \hat{\cal H} = \sum_{\bm{i}\not=\bm{j}} \left( {\hat c}^{\dagger}_{\bm{i}} {\hat F}^{\,}_{\bm{ij}} {\hat c}^{\,}_{\bm{j}} + \textrm{H.c.} \right) \,,\label{eq:controlf}
\end{equation}
where ${\hat c}^{\,}_{\bm{i}} ({\hat c}^{\dagger}_{\bm{i}})$ is a regular particle annihilation (creation) operator acting at position $\bm{i}$ of $\Lambda$. An appropriate choice of the operator-valued function ${\hat F}^{\,}_{\bm{ij}}$ can result in any possible $n$-body intersite term. For example, setting ${\hat F}^{\,}_{\bm{ij}}=1$ yields all possible hopping terms with equal amplitude, whereas ${\hat F}^{\,}_{\bm{ij}} = {\hat c}^{\,}_{\bm{i}} {\hat c}^{\dagger}_{\bm{j}}$ yields all possible density-density interactions with equal magnitude.

In the following, we restrict the ${\hat F}^{\,}_{\bm{ij}}$ to products of hole-density operators of the form $(1-\hat n)$ acting in the neighborhood of $\bm{i}$ and $\bm{j}$. Such terms emerge effectively from strong interactions~\cite{Fendley2003}. To see this, consider particles that hop around the lattice and at the same time interact with each other via a nearest-neighbor repulsion of strength $V$. In the infinite-$V$ limit, particles cannot occupy nearest-neighbor sites. This suggests that states containing nearest-neighboring particles can be removed from the Fock space. Accordingly, for any site $\bm{i}\in\Lambda$ we define the projected operator $\tilde{c}^{\dag}_{\bm{i}}$ by demanding that its action on any state is to create a particle at $\bm{i}$ if and only if this site and all of its nearest neighbors are empty. Formally,
\begin{equation}
\tilde{c}^{\dag}_{\bm{i}}:=\hat{c}^{\dag}_{\bm{i}}\prod_{\lbrace \bm{j} : |\bm{j}-\bm{i}|=1 \rbrace}\left(1-\hat{n}^{\,}_{\bm{j}}\right) \,,
\end{equation}
where ${\hat n}^{\,}_{\bm{j}} := {\hat c}^{\dagger}_{\bm{j}} {\hat c}^{\,}_{\bm{j}}$. If we allow particles to hop while obeying the above hard-core condition, then the system can be described by Eq.~\eqref{eq:controlf} with
\begin{equation}
 {\hat F}^{\textrm{HC}}_{\bm{ij}} = t^{\,}_{\bm{ij}} \prod_{\lbrace \bm{l} : |\bm{l}-\bm{i}|=1 \vee |\bm{l}-\bm{j}|=1 \rbrace}\left(1-\hat{n}^{\,}_{\bm{l}}\right) \,,\label{eq:hardsph}
\end{equation}
where $t^{\,}_{\bm{ij}}$ are generally complex-valued hopping amplitudes. The strong-repulsion limit can therefore be recast into a Hamiltonian of the form of Eq.~\eqref{eq:controlf}, with ${\hat F}^{\,}_{\bm{ij}}$ being a product of hole-density operators in the vicinity of $\bm{i}$ and $\bm{j}$. On two-dimensional lattices of corner- or edge-sharing triangles, the allowed states are exactly the configurations of classical HH models, widely studied in the context of glassiness. Due to this resemblance, we shall call the above Hamiltonians quantum hard-hexagon (QHH) models.

We now wish to reduce the above hardcore constraint to a new one that does not correspond to a density-density interaction. One motivation for doing so is that density-hopping terms of the form $\hat c^\dagger (1- \hat n) \hat c$ have been shown to arise in the description of optical lattice experiments, where trapped atoms are periodically driven using circularly polarized light or equivalent settings~\cite{Aidelsburger2014,Jotzu2014}. The theoretical treatment of relevant models describes how the driving, apart from affecting the preexisting hopping and density-density terms, introduces new frequency-dependent density-hopping terms~\cite{Bukov2014}.
% Instead of treating the latter in conjunction with other processes in a complicated setting, here we isolate them and study their properties in a simpler context.
The relevance of these terms in the experimental setups depends on the interaction strength between particles, which is in principle tunable. Nevertheless, their precise effect in that setting requires careful study, as it may very well be scrambled by other processes. As a first step, however, it is of intrinsic interest to isolate the kinetically constrained terms and determine their properties in a simpler context. As we shall show below, such terms can generate nontrivial behavior even on their own.

Notice that ${\hat F}^{\textrm{HC}}$ is just a product of hole-density operators at different sites. In any partially filled system, products of $(1-\hat n)$ operators that act on different sites are more likely to vanish than single $(1-\hat n)$ operators. We can therefore attempt to restrict the product in Eq.~\eqref{eq:hardsph} to only a few $(1-\hat n)$ operators and see whether this captures the same physics as the full product. To decide which terms to truncate, we draw inspiration from classical models of diffusion. As mentioned above, ${\hat F}^{\textrm{HC}}$ is the quantum counterpart of the HH model on the triangular lattice. In the HH model, each particle can be visualized as a hard disk of radius $1/2 < r < 1$, so that configurations with particles on nearest-neighboring sites are forbidden. If the disk radius is reduced to $\sqrt{3}/4 < r < 1/2$, then all particle configurations are a priori allowed. The particle motion, however, is still constrained if we assume that hard discs move from site to site along straight lines: On lattices of edge- or corner-sharing equilateral triangles, they cannot move past one another (see Fig.~\ref{fig:cartoon}). This is an example of a cooperative lattice-gas model~\cite{Jackle2002}.

A quantum analog of the kinetically constrained model described above can be straightforwardly constructed by discarding all factors in Eq.~\eqref{eq:hardsph} apart from those which pertain to common neighbors of $\bm{i}$ and $\bm{j}$. With this choice,
\begin{equation}
 {\hat F}^{\textrm{KC}}_{\bm{ij}} = t^{\,}_{\bm{ij}} \prod_{\lbrace \bm{l} : |\bm{l}-\bm{i}|=|\bm{l}-\bm{j}|=1 \rbrace}\left(1-\hat{n}^{\,}_{\bm{l}}\right) \,. \label{eq:kc}
\end{equation}
Examples of the resulting terms are pictorially represented in Fig.~\ref{fig:cartoon} for two lattices. Following the classical terminology, we shall call the processes represented by ${\hat F}^{\textrm{KC}}_{\bm{ij}}$ vacancy-assisted hoppings (VAH). Note that this type of constraint is not specific to lattices containing triangular plaquettes: Next-nearest neighbor hoppings on checkerboard or honeycomb lattices can be similarly constrained. Here we focus on the triangular and kagome lattices because all hoppings present in the models introduced below are constrained in the manner enforced by Eq.~\eqref{eq:kc}, and also because these models are known to yield the most robust fermionic FCI states.

\subsection{Example models}

We now ask whether the correlations induced by the vacancy-assisted hoppings described above can generate topological order. To answer this, we shall impose this kinetic constraint to tight-binding models with topologically nontrivial bands. We choose $t^{\,}_{\bm{ij}}$ that correspond to two thoroughly studied FCI models, diagonalize the constrained Hamiltonian exactly on finite periodic clusters and look for signatures of topologically nontrivial states.

\begin{figure}[t]
\begin{center}
 \includegraphics[width=\columnwidth]{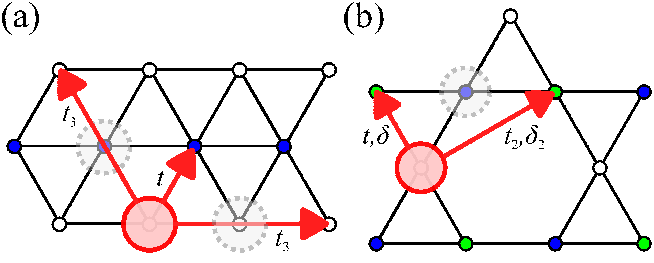}
 \end{center}
 \caption{(Color online) Illustration of vacancy-assisted hopping models on (a) the triangular lattice, with nearest and third-nearest neighbor hoppings as defined in Refs.~\onlinecite{Venderbos2011a,Kourtis2012a}, and (b) the kagome lattice, with nearest and second-nearest neighbor hoppings as defined in Ref.~\onlinecite{Tang2011}. Solid and dashed circles denote particles and vacancies, respectively. The hoppings (arrows) are only allowed if a vacancy is located at the position shown. Note that some of the hoppings are imaginary; see Eqs.~\eqref{eq:tri} and~\eqref{eq:kag}.}
 \label{fig:cartoon}
\end{figure}

Tight-binding Hamiltonians are represented as ${\cal H}:=\sum_{\bm{k}} \psi^\dagger_{\bm{k}} H^{\,}_{\bm{k}} \psi^{\,}_{\bm{k}}$, where $\psi^{\,}_{\bm{k}}\equiv (c^{\,}_{1,\bm{k}}, c^{\,}_{2,\bm{k}}, ... , c^{\,}_{n,\bm{k}})^{\mathsf{T}}$ is the spinor of annihilation operators on each of the $n$ sublattices of the lattice model at momentum $\bm{k}$. The triangular-lattice model introduced in Refs.~\onlinecite{Venderbos2011a,Kourtis2012a} is given by
\begin{subequations}
\begin{align}
H^{\bigtriangleup}_{\bm{k}} :=&{\ } \bm{g}^{\,}_{\bm{k}} \cdot \boldsymbol\tau \,,\\
g^{\,}_{0,\bm{k}} =&{\ } 2t^{\,}_3 \sum_{j=1}^3 \cos(2{\bm{k}}\cdot\bm{a}^{\,}_i)\,,\\
g^{\,}_{i,\bm{k}} =&{\ } 2t \cos({\bm{k}}\cdot\bm{a}^{\,}_i), \qquad i=1,2,3,
\end{align}\label{eq:tri}%
\end{subequations}
where $\bm{a}^{\,}_1 = (1/2, -\sqrt{3}/2)^{\mathsf{T}}$, $\bm{a}^{\,}_2 = (1/2, \sqrt{3}/2)^{\mathsf{T}}$, $\bm{a}^{\,}_3 = -( \bm{a}^{\,}_1+\bm{a}^{\,}_2 )$ and $\boldsymbol\tau \equiv (\tau^{\,}_0,\tau^{\,}_1,\tau^{\,}_2,\tau^{\,}_3)$ is the vector of Pauli matrices including the $2\times2$ unit matrix as $\tau^{\,}_0$. The kagome-lattice model introduced in Ref.~\onlinecite{Tang2011} is defined as
\begin{subequations}
\begin{align}
H^{\davidsstar}_{\bm{k}} :=&{\ } \bm{d}^{\,}_{\bm{k}} \cdot \boldsymbol\lambda + \tilde{\bm{d}}^{\,}_{\bm{k}} \cdot \tilde{\boldsymbol\lambda} \,,\\
d^{\,}_{i,\bm{k}} =&{\ } -2t \cos({\bm{k}}\cdot\bm{a}^{\,}_i) + 2t^{\,}_2 \cos({\bm{k}}\cdot\bm{b}^{\,}_i), \ i=1,2,3,\\
\tilde d^{\,}_{i,\bm{k}} =&{\ } -2\delta \cos({\bm{k}}\cdot\bm{a}^{\,}_i) + 2\delta^{\,}_2 \cos({\bm{k}}\cdot\bm{b}^{\,}_i), \ i=1,2,3,
\end{align}\label{eq:kag}%
\end{subequations}
where $\bm{a}^{\,}_1 = (1,0)^{\mathsf{T}}$, $\bm{a}^{\,}_2 = (1/2,\sqrt{3}/2)^{\mathsf{T}}$, $\bm{a}^{\,}_3 = \bm{a}^{\,}_2-\bm{a}^{\,}_1$, $\bm{b}^{\,}_1 = \bm{a}^{\,}_2+\bm{a}^{\,}_3$, $\bm{b}^{\,}_2 = \bm{a}^{\,}_3-\bm{a}^{\,}_1$, $\bm{b}^{\,}_3 = \bm{a}^{\,}_1+\bm{a}^{\,}_2$ and $\boldsymbol\lambda \equiv (\lambda^{\,}_1,\lambda^{\,}_2,\lambda^{\,}_3)$, $\tilde{\boldsymbol\lambda} \equiv (\tilde\lambda^{\,}_1,\tilde\lambda^{\,}_2,\tilde\lambda^{\,}_3)$ are vectors of the Gell-Mann matrices
\begin{subequations}
\begin{align}
 \lambda^{\,}_1 &= \begin{pmatrix}
                   0 & 1 & 0 \\
                   1 & 0 & 0 \\
                   0 & 0 & 0 
                  \end{pmatrix},
 \lambda^{\,}_2 = \begin{pmatrix}
                   0 & 0 & 1 \\
                   0 & 0 & 0 \\
                   1 & 0 & 0 
                  \end{pmatrix},
 \lambda^{\,}_3 = \begin{pmatrix}
                   0 & 0 & 0 \\
                   0 & 0 & 1 \\
                   0 & 1 & 0 
                  \end{pmatrix}, \\
 \tilde\lambda^{\,}_1 &= \begin{pmatrix}
                   0 & \mathrm{i} & 0 \\
                   -\mathrm{i} & 0 & 0 \\
                   0 & 0 & 0 
                  \end{pmatrix},
 \tilde\lambda^{\,}_2 = \begin{pmatrix}
                   0 & 0 & -\mathrm{i} \\
                   0 & 0 & 0 \\
                   \mathrm{i} & 0 & 0 
                  \end{pmatrix},
 \tilde\lambda^{\,}_3 = \begin{pmatrix}
                   0 & 0 & 0 \\
                   0 & 0 & \mathrm{i} \\
                   0 & -\mathrm{i} & 0 
                  \end{pmatrix}.
\end{align}
\end{subequations}
We set $t^{\,}_3/t = 0.19$ for the triangular lattice and $t^{\,}_2/t=0.3, \delta/t=0.28, \delta^{\,}_2/t=0.2$ for the kagome lattice. These values of the hopping terms are chosen to yield relatively flat lowest bands with Chern number $C=-1$. We have verified that all our results remain valid in a finite range around the values chosen here. We focus on average particle densities $\rho=1/6$ for the triangular lattice and $\rho=1/9$ for the kagome lattice, at which the lowest band of the pure hopping counterparts of the models we study here would be at filling $\nu=1/3$. We then impose the constraint encapsulated in Eq.~\eqref{eq:kc}. The resulting vacancy-assisted hopping models are sketched in Fig.~\ref{fig:cartoon}.

\begin{figure}[t]
\begin{center}
 \includegraphics[width=\columnwidth]{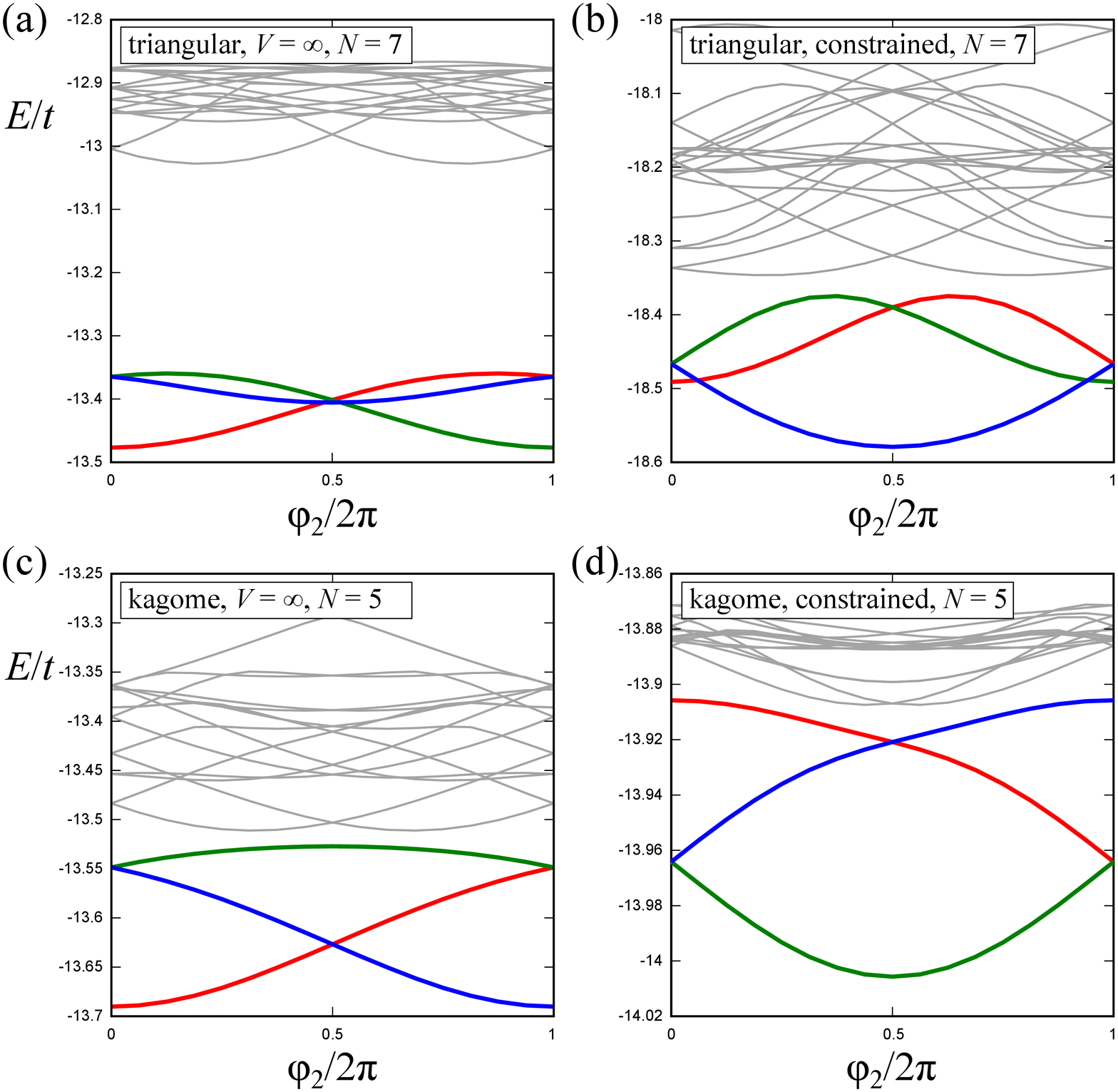}
 \end{center}
 \caption{(Color online) Flow of eigenvalues under insertion of magnetic flux corresponding to a phase $\varphi^{}_2$ through one of the handles of the toroidal system for (a),(b) the triangular and (c),(d) the kagome lattice. (a), (c) correspond to QHH models (i.e., ${\hat F}={\hat F}^{\textrm{HC}}$) and (b), (d) to VAH models (i.e., ${\hat F}={\hat F}^{\textrm{KC}}$).}
 \label{fig:flow}
\end{figure}

\section{Results}\label{sec:results}

FQH-like topological order at $\nu=1/3$ manifests itself in the many-body energy spectrum as spectral flow, which is the exchange of the threefold quasidegenerate many-body ground-state levels upon insertion of a flux quantum through one of the handles of the toroidal (i.e., periodic) system. This spectral property is shown in Fig.~\ref{fig:flow}. To establish consistency with previous results~\cite{Kourtis2013a}, Fig.~\ref{fig:flow}(a) shows the spectral flow of the QHH version of the triangular-lattice model, where simultaneous occupation of nearest-neighboring sites is prohibited but hoppings are unconstrained. In Fig.~\ref{fig:flow}(c) we show that the same physics occurs in the kagome-lattice model. Apart from the characteristic FQH features in the energy spectra, the ground states obtained have a very precisely quantized Hall conductivity $\sigma^{}_H = e^2/(3h)$.

\begin{figure}[t]
\begin{center}
 \includegraphics[width=\columnwidth]{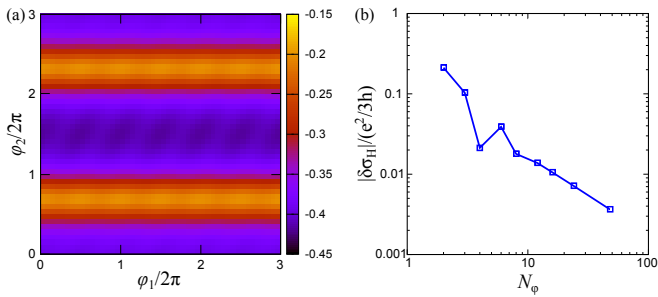}
 \end{center}
 \caption{(Color online) (a) Berry curvature of the triangular-lattice model on a $42$-site cluster with $N=7$ as a function of the magnetic fluxes $\varphi_1$ and $\varphi_2$ threading the two handles of the toroidal system; (b) error in the quantization of $\sigma^{}_H$ as a function of the number of subintervals in the partition of the Brillouin zone of fluxes. The unidirectionality of the stripes along $\varphi_1$ is merely an artifact of the choice of the alignment of the two-site unit cell of the model [see Fig.~\ref{fig:cartoon}(a)].}
 \label{fig:berry}
\end{figure}

We now reduce the HH constraint of Eq.~\eqref{eq:hardsph} to the kinetic constraint of Eq.~\eqref{eq:kc}, as outlined above. We notice that the energy spectra, shown on the right of Fig.~\ref{fig:flow}, are qualitatively equivalent. The symmetry sectors in which the quasidegenerate ground states reside are the same as in the case of hardcore interactions and can be predicted by methods already in use for finite FQH and FCI systems~\cite{Regnault2011,Bernevig2012}. More importantly, the ground states of the VAH models have nontrivial topological characteristics. Their many-body Berry curvature is a smooth function that integrates to a precisely quantized Hall conductivity, even for the finite systems considered here. An example of this quantization is presented in Fig.~\ref{fig:berry} for the triangular-lattice model. Equivalent results can be obtained for the kagome-lattice model as well.

\begin{figure}[t]
\begin{center}
 \includegraphics[width=\columnwidth]{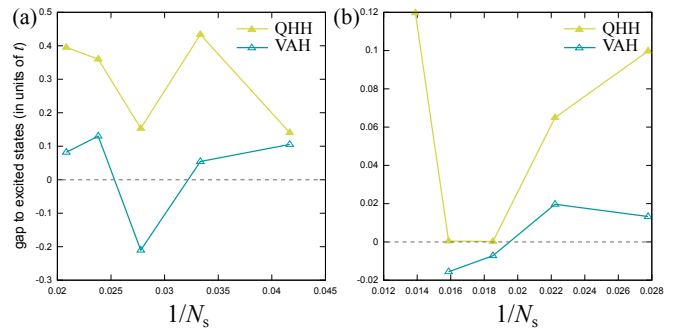}
 \end{center}
 \caption{(Color online) Gap to excited states at zero flux as a function of inverse number of lattice sites $N^{}_s$ for QHH (solid triangles) and VAH (empty triangles) on (a) the triangular-lattice model and (b) the kagome-lattice model. The value of the gap is the energy difference between the highest in energy quasidegenerate FCI level and the lowest level that does not belong to the FCI state manifold.}
 \label{fig:fsize}
\end{figure}

The FQH-like ground states of the VAH models are generically less gapped than those of their QHH counterparts, as can be seen in Fig.~\ref{fig:fsize}. Nevertheless, there is a finite volume in parameter space in which one obtains gapped FQH-like ground states even in the VAH models for most clusters. Finite-size effects --- such as the noticeable energy separation between quasidegenerate ground-state levels --- are sizable and do not allow for a definitive answer as to whether there is topological order in the ground state of the VAH models in the thermodynamic limit. We note, however, that the gaps of the VAH models do seem to follow the same trend as those of their QHH counterparts, which remain well gapped up to the largest accessible system sizes (see also Ref.~\onlinecite{Kourtis2013a}). The finite-size effects observed in our calculations are similar to the ones observed in previous, more detailed studies of the dependence of FCI state energies on cluster size and aspect ratio, which concluded that these effects are indeed irrelevant to the underlying physics~\cite{Lauchli2012}. Furthermore, we find no signatures of a charge instability, so the only evident competitor for the ground state is a compressible metallic state. The latter is clearly disfavored according to previous detailed studies of FCIs in the models defined by Eqs.~\eqref{eq:tri} and~\eqref{eq:kag} in the parameter regimes chosen here~\cite{Wu2012,Kourtis2012a}.

\section{Conclusion}\label{sec:conclusions}

Regardless of the detailed energetics of the models derived above, the key conclusion of this work is that kinetic constraints in partially filled Chern bands can generate topologically ordered states.
% Given the relevance to current optical-lattice setups~\cite{Aidelsburger2014,Jotzu2014,Bukov2014}, this may have direct consequences on the realization of topologically ordered cold atoms. Moreover, we have argued that 
This can be intuitively understood by noting that kinetic constraints can be seen as an approximate generalization of hardcore nearest-neighbor repulsion. The ground states of the latter can be adiabatically connected to FCI states, which in turn means that correlations crucial for FQH-like states can be generated by kinetic constraints alone. This may be a stepping stone for further intuition into the microscopics of the FQH effect itself. More hints in this direction can be found in recent work~\cite{Lee2013}, where the intimate relation between FQH and FCI states gives rise to similar density-hopping terms. Finally, we briefly comment on choosing to introduce fermionic models in this work. The reason for our choice is that the $\nu=1/3$ fermionic FCI state is the most robust state --- and therefore the easiest one to study --- that requires short-range interactions. If one is interested in modeling current experiments precisely, it would be more meaningful to pursue bosonic FCI states. However, the most robust bosonic FCI state at $\nu=1/2$ requires only an on-site interaction, whereas the $\nu=1/4$ state requires longer-range repulsion, which would consequently give rise to less intuitive kinetic constraints. Since here we are interested in whether kinetic constraints can in principle induce topological order, we have focused on the simpler case of the $\nu=1/3$ fermionic FCI. Nevertheless, similar tendencies towards topological ordering are to be expected when suitable kinetic constraints are introduced in bosonic versions of Haldane-like models~\cite{Wang2011}, with potential repercussions on the realization of topologically ordered cold-atom systems~\cite{Aidelsburger2014,Jotzu2014,Bukov2014}.

% As a closing remark, we mention another tantalizing outlook on our results, which is the possibility of establishing a connection with the statistical mechanics of classical systems. The classical counterparts of the QHH and VAH models introduced here were invented to model dynamics in glassy systems. Typically, the models pertaining to glassiness have long thermalization times and support correspondingly long-lived excitations. It would be of great interest to investigate whether some of this glassiness also applies to fractionalized quasiparticle excitations, which could then be more easily manipulated for functional purposes.

\begin{acknowledgments}
This work was supported in part by Engineering and Physical Sciences Research Council Grant No. EP/G049394/1, the Helmholtz Virtual Institute ``New States of Matter and Their Excitations'' and the EPSRC NetworkPlus on ``Emergence and Physics far from Equilibrium''. S.K. acknowledges financial support by the ICAM Branch Contributions. The authors are grateful to M.~Bukov, C.~Chamon, N.R.~Cooper, M.~Daghofer, A.G.~Grushin, C.~Mudry, T.~Neupert, and J.K.~Pachos for stimulating discussions. 
\end{acknowledgments}

% % % Phys. Rev. Lett. 87, 120405 (2001)
% % % Phys. Rev. B 71, 121303(R) (2005)
% % % Phys. Rev. Lett. 110, 185301 (2013)
% % % 
% % % Constraints:
% % % Phys. Rev. Lett. 108, 045306 (2012)

% \bibliography{/home/stefanos/Documents/phys/bibtex/library}
\bibliographystyle{apsrev4-1}
% \bibliography{bib}

%

\end{document}